\documentclass{article}
\usepackage[utf8]{inputenc}
\usepackage{graphicx}
\usepackage[section]{placeins}
\usepackage{indentfirst}
\usepackage{float}
\usepackage[table,xcdraw]{xcolor}
\newcommand{\gd}{\color{black}}
\newcommand{\kr}{\color{black}}
\newcommand{\djj}{\color{black}}
\usepackage[normalem]{ulem}

\title{Decoding planetary surfaces by counting cracks}
\author{S. Silver*, K. Reg\H{o}s*, D.J. Jerolmack**, and G. Domokos**\\
**these authors contributed equally. \\
***corresponding authors}
\date{\today}

\begin{document}

\maketitle

\section*{Abstract}
{\djj{Planets are often covered with thin cracked shells. From mud films to lithospheres of rock or ice \cite{El-Maarry2014, Montigny2022, Mangold2005, Smrekar2002, Rudolph2009}, fracture networks form two-dimensional (2D) tesselations of convex polygons whose geometry encodes their genesis \cite{Domokos2020, Domokos2022}. Here we chart the geometry of 2D fracture mosaics across the solar system, and decode their formative conditions using a new dynamical crack model. We show that mosaics can be projected onto a Symbolic Ternary Diagram, where the relative proportions of ``T'', ``X'' and ``Y'' junctions are uniquely related to contributions from distinct modes of fracture.}}
Most planetary mosaics cluster in a region associated with hierarchical fracture networks, where sequential cracking favors formation of T junctions \cite{Goehring, Ma2019}. Exceptions to this rule may betray the presence of water. Europa's fracture networks \cite{Greeley1998} stand apart due to the predominance of X junctions; this is a special feature of ice, where healing of cracks by refreezing of water allows new fractures to overprint older ones \cite{Culberg2018}. Several fracture networks on Mars appear as outliers due the high proportion of Y junctions. These patterns -- previously interpreted as ancient mudcracks and frozen polar terrain, based on geological evidence --  are consistent with the twisting of crack junctions by cyclic volume change \cite{Goehring}. Our findings suggest that counting cracks could aid in the identification of other water-influenced planetary environments.

\section*{Main}

Things fall apart. From the moment solid rock or ice is formed, stresses inexorably fragment this material into ever more intricate fracture mosaics; witness the cracked outer shells of planetary bodies across the solar system. 
At the smallest scales, mm-thin layers of drying mud give rise to cm-size mudcracks. At the largest scales, the lithospheres of gas-giant moons
are torn asunder by tidal forces. Surface fracture patterns are at the center of some of the most important questions in planetary evolution: did martian mudcracks require a prolonged hydrologic cycle to form \cite{McKeown2013, rapin2023sustained, McKinnon2016,Trowbridge2016}?; do fractures on Venus signal some kind of plate tectonics? \cite{Byrne2021}; and are the double-ridge cracks on Europa a portal to a subsurface ocean of water that harbors life \cite{Howell2020}? These questions motivate this paper.   

Numerous studies on Earth have demonstrated how distinct fracture patterns can arise from differences in formative stresses or material composition \cite{Goehring, Ma2019, Bahat2005}. On other planets, however, we are unfortunately lacking the critical information needed to make such linkages.
Images are often the primary -- and sometimes the only -- observations that we have of planetary fracture mosaics. Numerical models have been useful for assessing the plausibility of competing hypotheses for pattern formation, but such models can be difficult to constrain. Can we use the geometry of crack networks to provide better constraints? There are two challenges to doing so: (i) different formative processes can produce the same apparent pattern; 
and (ii) the diversity of crack patterns is large and growing, as our catalog of planetary observations increases. A recent study, however, has shown that these challenges can be overcome -- or at least greatly simplified -- by examining fracture patterns through the lens of convex mosaics \cite{Domokos2020}. The idea is that natural fracture cells are well approximated as convex polygons, and that this places strong geometric constraints on the structure of fracture mosaics \cite{Domokos2020}. As an example, consider the special case of two dimensional (2D), regular polygons: only triangles, rectangles and hexagons tessellate. Although real fracture cells are not regular polygons, 
their average geometric properties are bounded by these end-member shapes \cite{Domokos2020}. More, rectangular and hexagonal cell patterns were shown to be attractors associated with distinct stress fields. Deviatoric stresses form{\gd{, on average,}} rectangular cells (the Platonic attractor), while isotropic stresses or cyclic volume change make{\gd{, on average,}}  hexagonal cells (the Voronoi attractor) \cite{Domokos2020}. 

\djj{In this study we chart the geometry of fracture networks on planetary surfaces across the solar system. We introduce an intuitive representation of this geometry in the form of a Symbolic Ternary Diagram, that plots the relative proportion of T, X and Y junctions. We find that most planetary surfaces represent highly-evolved hierarchical fracture networks dominated by T junctions, that have not experienced rejuvenation or resurfacing. Important exceptions are: (i) frozen polar deposits and putative mudcracks on Mars -- both dominated by Y junctions -- that likely required many cycles of contraction and expansion; and (ii) Europa's ice cracks dominated by X junctions, which we suggest anneal by refreezing at rates comparable to new fracture formation. We use a new dynamical crack model \cite{Balint2023} to demonstrate that the three distinct clusters of network geometry correspond to distinctly different formative mechanisms. More, we show how this dynamical model can be used to trace the entire trajectory of fracture network evolution, both forward and backward in time.  }
{\djj{This work demonstrates a promising avenue for measuring and interpreting planetary data using geometry.}}

\section*{Theory}
\subsection*{Geometric description of mosaics}
\begin{figure} [ht]
    \includegraphics[width=12cm]{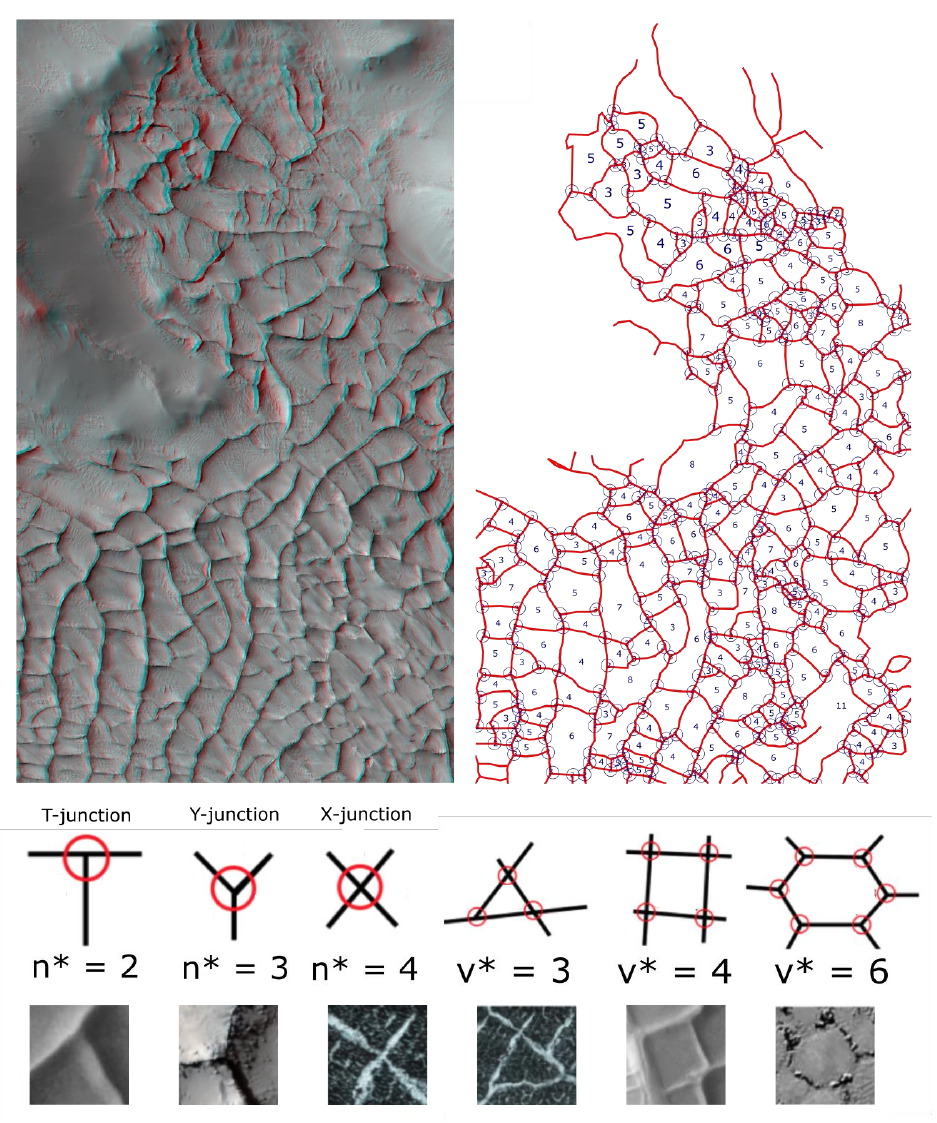}
    \centering
    \caption{\textbf{Counting crack networks as convex mosaics.} Top left: image of Gordii Dorsum Ridge on Mars. Top right: skeletonized network with values of $n^*$ and $v^*$ superimposed. Bottom: Examples of nodes and cells, illustrated and sampled from real planetary images.}\label{fig:1}
\end{figure}

\begin{figure} [ht]
    \includegraphics[width=12cm]{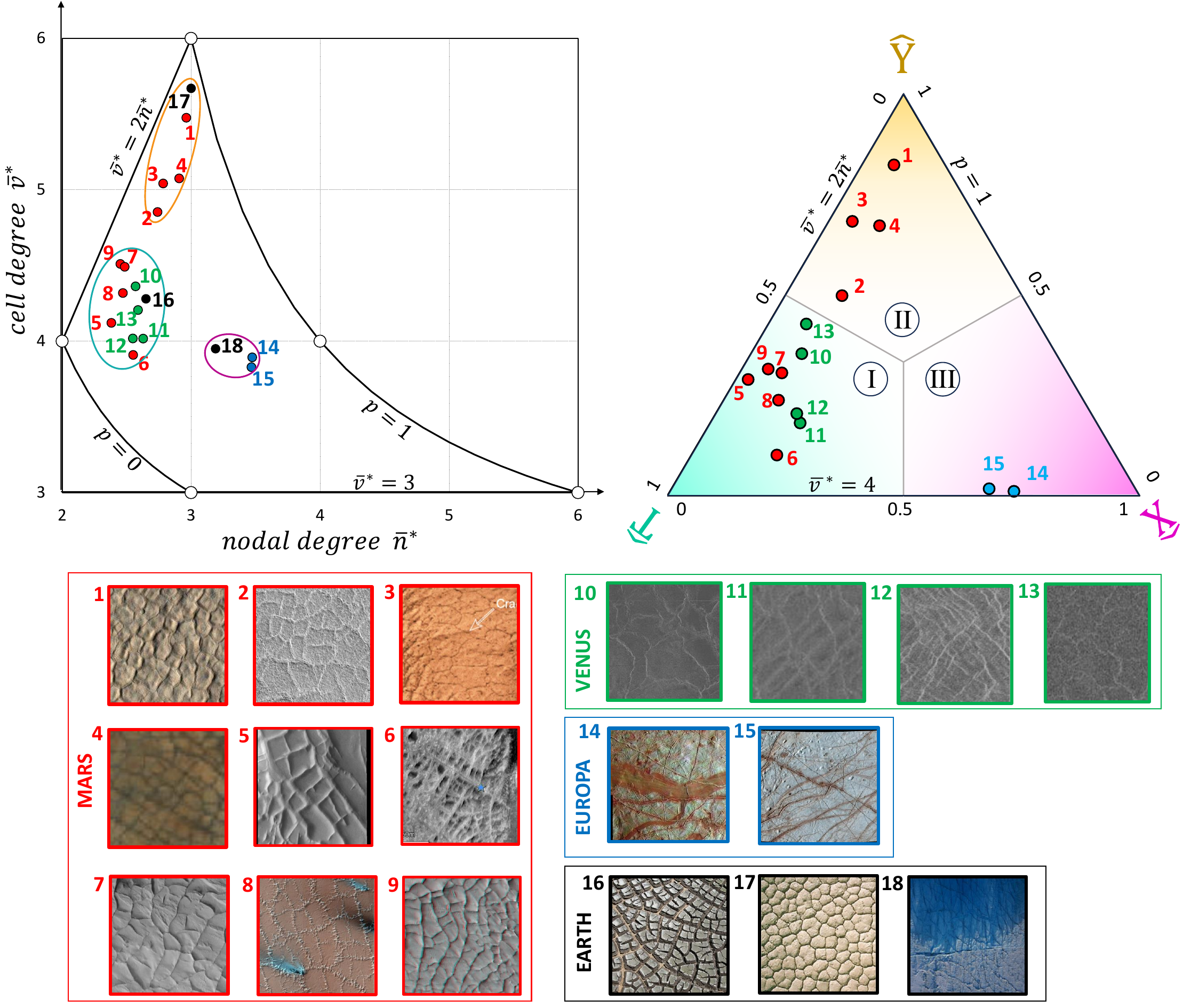}
    \centering
    \caption{\textbf{Planetary fracture mosaics form three clusters.} Top Left: Static view of the planetary Symbolic Plane. Data from planetary fracture mosaics break into three clusters, corresponding to qualitatively distinct fracture mosaics. A representative Earth example of each type is also shown. Images correspond to plotted mosaics. 
    Top right: The Symbolic Ternary Diagram for planetary fracture mosaics (Earth mosaics are excluded). The diagram can be partitioned into three regions, corresponding to the dominance of distinct node types: I (T nodes), II (Y nodes) and III (X nodes). The three clusters observed in the Symbolic Plane (Fig. 2) separate into the three partitions here. The majority of observed mosaics reside in partition I. Mosaics in partition II are only found on Mars in materials suspected to experience cyclic volume change. Mosaics in partition III only appear on Europa, likely due to crack healing. Numbers in both plots correspond to images at bottom, and data in Table \ref{table:1}.
    }
    \label{fig:2}
\end{figure}

Starting with the assumption that cells within a fracture mosaic are well approximated as convex polygons, the pattern can be completely described statistically using three quantities that are derived from integers \cite{Domokos2020, Domokos2022}; that is, by counting cracks and their intersections (Fig. \ref{fig:1}). The cell degree {\gd{of an infinite mosaic}}, $\bar{v}^{\star}$, is the average number of {\gd{sharp}} vertices, $v^{\star}$, per cell in the mosaic. The nodal degree of {\gd{an infinite}} mosaic, $\bar{n}^{\star}$, is the average number of {\gd{sharp}} cell vertices that meet at an intersection, $n^{\star}$. Together these two quantities define the Symbolic Plane, $[\bar{n}^{\star}, \bar{v}^{\star}]$ (Fig. \ref{fig:2}; Fig. S3).
{\gd{We distinguish finite mosaics by the subscript 1 and we call the point $M_1(\bar{n}_1^{\star}, \bar{v}_1^{\star})$ the \emph{static estimate} for the location of the mosaic in the Symbolic Plane.}} The third parameter is the regularity, $0 \leq p = N_R/(N_I + N_R) \leq 1 $, where $N_R$ is the number of regular nodes (defined here as X or Y junctions) and $N_I$ is the number of irregular nodes (defined here as T junctions; other studies have defined irregularity differently \cite{ternary1}). For infinite mosaics the three parameters are exactly related \cite{Domokos2020} via
\begin{equation}\label{eq:symbolic}
    \bar{v}^{\star} =\frac{ 2 \bar{n}^{\star}}{\bar{n}^{\star} -p -1};
\end{equation}
this relation is approximate for finite mosaics. The universal domain of possible {\gd{convex}} fracture patterns in the Symbolic Plane -- that is, patterns permitted by geometry -- is determined by these quantities through the theoretical relations
{\gd{$0 \leq p \leq 1$ and $3 \leq \bar v^{\star}\leq 2 \bar n^{\star}$ \cite{Domokos2020} (Fig. \ref{fig:2}; Fig. S3)}}. 

Fracture mechanics appears to select a subrange of the mathematically possible space. {\djj{In particular, we recognize three end-member patterns that approximately bound the domain of observed fracture patterns on Earth: Regular, Irregular, and Voronoi mosaics \cite{Domokos2020} made entirely of X, T, and Y junctions, respectively (Fig. \ref{fig:2}). }}
{\gd{Defining the relative frequencies of each as $\hat X=X/(X+Y+T),$ $\hat Y=Y/(X+Y+T)$
and $\hat T=1-\hat X-\hat Y$, we can represent these mosaics on a Symbolic Ternary Diagram (Figure \ref{fig:2}). (This plot is reminiscent, but essentially different from, the ternary plot used in the purely combinatorial theory \cite{ternary1, ternary}).} \djj{We refer to any infinite mosaic with nodes composed only of T, Y and X junctions as a \emph{TYX mosaic}. For these idealized mosaics, there is a direct mapping from the Symbolic Plane to the Symbolic Ternary Diagram via equation (\ref{eq:symbolic}) :}
\begin{equation}\label{eq:dynamic}
    \bar{n}_2^{\star}=2+\hat Y + 2\hat X, \quad  p=\hat X+ \hat Y, \quad \bar{v}_2^{\star}=4+\frac{2\hat Y}{1 + \hat X},
\end{equation}
(see Supplementary Data Section 4 for more detail).  For TYX mosaics the two representations are equivalent, thus we have $M_2(\bar{n}_2^{\star}, \bar{v}_2^{\star})\equiv M_1(\bar{n}_1^{\star}, \bar{v}_1^{\star})$. \djj{Non-TYX mosaics are either finite in size, or contain other junction types beyond T, Y and X (or both). For these mosaics, the coordinates $[\bar{n}_1^{\star}, \bar{v}_1^{\star}]$ in the Symbolic Plane are not equivalent to the values $[\bar{n}_2^{\star}, \bar{v}_2^{\star}]$ inferred from the Symbolic Ternary Diagram. The difference between these two representations can be understood by modeling the dynamical evolution of fracture mosaics. }



\subsection*{Dynamical fracture model}
Distinct modes of fracture can be translated into geometric rules, which allows construction of a dynamical theory for the evolution of fracture mosaics.
A simple model that encodes only primary fracture as random cuts, and secondary fracture as binary breaking of cells, was found to quantitatively reproduce commonly observed Irregular fracture patterns \cite{Domokos2020}. This model can be generalized to simulate a broader class of fracture mosaics, by adding two physically-motivated rules. First is the T $\rightarrow$ Y conversion of nodes, which is known to result from cyclic volume change as observed for mudcracks under wetting-drying cycles \cite{Goehring2010}. The second is the overprinting of cracks by subsequent generations of (secondary) fracture, which is prevalent in ice due to healing from refreezing and sintering \cite{murdza2022rapid, murdza2023behavior, demmenie2022scratch, colbeck1986theory, schulson2016restoration}. We construct a {\gd{continuous time}} fracture model \cite{Balint2023} {\gd{driven by}} two local, {\gd{discrete}} steps: step $R_{0}^m$ creates secondary cracks with an intensity factor $\lambda_0$, and step $R_1$ is associated with the T $\rightarrow$ Y transition with an intensity factor $\lambda_1$ (Fig. \ref{fig:3}; Fig S4). The ratio $\mu = \lambda_0/ \lambda_1$ represents the relative rates of these two processes; $\mu = 0$ corresponds to no secondary fracturing. Each secondary crack traverses a number of cells represented by the parameter $m$, where $m = 0$ corresponds to the limit of cracks that span only one cell (binary breakup; Fig. \ref{fig:3}, Fig. S4).

{\gd{This}} dynamical fracture model creates only T, Y and X nodes.} \djj{In the limit of infinite time, every mosaic will evolve to a TYX mosaic regardless of its initial conditions. More, each physically admissible parameter pair [$m$, $\mu$] produces a globally attractive and unique fixed point $[\bar{n}^{\star}, \bar{v}^{\star}]$. In other words, if a mosaic has reached maturity then the model parameters -- and hence, formative fracture processes -- can be inferred from the geometry of the network. Although natural fracture mosaics are not likely to strictly meet these assumptions, they do not need to in order to relate observed fracture geometry to formative conditions. The dynamical model reveals that, for fixed model parameters $m$ and $\mu$, the fracture network evolves rapidly toward constant values of $\hat X, \hat Y$ and $\hat T$ (Supplementary Data Sections 4.4 and 5.4.). In practice, this means that the observed location of a fracture mosaic in the Symbolic Ternary Diagram (Fig. 2) corresponds to the expected fixed point in the dynamical fracture model. This allows us to determine the dynamical estimate $M_2(\bar{n}_2^{\star}, \bar{v}_2^{\star})$ using equation \ref{eq:dynamic}. Any differences between this dynamical estimate and the static estimate $M_1(\bar{n}_1^{\star}, \bar{v}_1^{\star})$ reflect the distance that an observed mosaic is from a (infinite) TYX mosaic (Fig. 3). }
\djj{Using the dynamical theory, the location $M_2$ of the attractor allows us to compute the system parameters $\mu,m$. To do so, we require that the model trajectory satisfy the following constraints: (i) it passes through the current location of the mosaic $M_1(\bar{n}_1^{\star}, \bar{v}_1^{\star})$ (the static estimate); (ii) it terminates at $M_2(\bar{n}_2^{\star}, \bar{v}_2^{\star})$ (the dynamic estimate); and (iii) it can be traced backward in time to a mosaic that we regard as a \emph{reasonable initial condition} (see Supplementary Data Section 4). Examples of such trajectories are shown in Fig. \ref{fig:4} (see Figs. S5 and S11 for more detail). }
{\djj{We use these principles below to test and confirm that all planetary mosaics analyzed below are mature, and can be used to infer fracture processes [Methods]. We encountered only one planetary mosaic that failed the above tests and therefore violated the maturity assumption. This failure resulted from strong boundary effects (Supplementary Data Section 5.1), and the mosaic was not included in our analysis below. This indicates that mosaics violating our assumptions may be rare, but they do indeed exist.  }}



%


{\djj{The $\mu=1$ and $m=1$ lines divide the Symbolic Plane into quadrants ({\gd{marked $A,B,C$ and $D$ in Fig. \ref{fig:3}}}), that delineate the dominance of distinct fracture processes.}} Quadrant A ($m < 1$, $\mu < 1$) contains the Irregular Limit, and corresponds to mosaics that form with little to no node-twisting or overprinting. Mosaics in this quadrant represent the predominance of binary breakup over all other processes. {\djj{The classic Gilbert model for crack growth, in which random line segments grow until they meet other lines, produces mosaics corresponding to this limit \cite{roy2022combinatorial}}}. A terrestrial example is thoroughly dried mud containing shrink/swell clays; these mosaics exhibit hierarchical crack patterns that arise from sequential secondary fracture \cite{Bohn2004, Ma2019}, and are composed predominantly of T nodes (Fig. \ref{fig:2}). Quadrant B ($m < 1$, $\mu >1$) contains the Voronoi Limit, and corresponds to mosaics dominated by node twisting with negligible overprinting (Fig. 3). Mosaics in this quadrant contain primarily Y junctions, that we attribute to cyclic volume change. The canonical terrestrial example is hexagonal mudcracks \cite{Domokos2020}, formed by repeated cycles of wetting and drying \cite{Goehring2010} (Fig. \ref{fig:2}). Quadrant C ($m > 1$, $\mu < 1$) contains the Regular Limit, and represents mosaics dominated by X junctions that result from significant overprinting and negligible node twisting (Fig. 3). We associate this quadrant with crack healing in fractured ice systems --- such as the West Greenland ice sheet, where it is known that healing of cracks by refreezing of water occurs \cite{Chudley2019} (Fig. \ref{fig:2}). Quadrant D ($m > 1$, $\mu > 1$) corresponds to the coexistence of node twisting and overprinting, which would produce networks with significant proportions of Y and X junctions, respectively. We have not found such mosaics in nature.

\begin{table}[]
    \centering
    \begin{tabular}{clrrrr}
\multicolumn{1}{r}{}      & Name & $\bar{n}^*$ & $\bar{v}^*$ & $\hat X$ & $\hat Y$  \\
{\color[HTML]{FE0000} 1}  & {\color[HTML]{FE0000} Amazonis   Planitia}     & {\color[HTML]{FE0000} 2,96} & {\color[HTML]{FE0000} 5,48} & {\color[HTML]{FE0000} 0,068}  & {\color[HTML]{FE0000} 0,825} \\
{\color[HTML]{FE0000} 2}  & {\color[HTML]{FE0000} Northern   Plains}       & {\color[HTML]{FE0000} 2,74} & {\color[HTML]{FE0000} 4,85} & {\color[HTML]{FE0000} 0,121}  & {\color[HTML]{FE0000} 0,498} \\
{\color[HTML]{FE0000} 3}  & {\color[HTML]{FE0000} Utopia   Planitia}       & {\color[HTML]{FE0000} 2,78} & {\color[HTML]{FE0000} 5,04} & {\color[HTML]{FE0000} 0,050} & {\color[HTML]{FE0000} 0,683} \\
{\color[HTML]{FE0000} 4}  & {\color[HTML]{FE0000} Mawrth Vallis} & {\color[HTML]{FE0000} 2,91} & {\color[HTML]{FE0000} 5,07} & {\color[HTML]{FE0000} 0,113}  & {\color[HTML]{FE0000} 0,674} \\
{\color[HTML]{FE0000} 5}  & {\color[HTML]{FE0000} Angustus Labyrinthus}    & {\color[HTML]{FE0000} 2,38} & {\color[HTML]{FE0000} 4,12} & {\color[HTML]{FE0000} 0,027}  & {\color[HTML]{FE0000} 0,293} \\
{\color[HTML]{FE0000} 6}  & {\color[HTML]{FE0000} Gale   Crater}           & {\color[HTML]{FE0000} 2,56} & {\color[HTML]{FE0000} 3,91} & {\color[HTML]{FE0000} 0,187}  & {\color[HTML]{FE0000} 0,104} \\
{\color[HTML]{FE0000} 7}  & {\color[HTML]{FE0000} Exhumed   Ridges}        & {\color[HTML]{FE0000} 2,48} & {\color[HTML]{FE0000} 4,49} & {\color[HTML]{FE0000} 0,090}  & {\color[HTML]{FE0000} 0,306} \\
{\color[HTML]{FE0000} 8}  & {\color[HTML]{FE0000} Sublimation   Polygons}  & {\color[HTML]{FE0000} 2,47} & {\color[HTML]{FE0000} 4,32} & {\color[HTML]{FE0000} 0,116}  & {\color[HTML]{FE0000} 0,238} \\
{\color[HTML]{FE0000} 9}  & {\color[HTML]{FE0000} Gordii Dorsum}           & {\color[HTML]{FE0000} 2,45} & {\color[HTML]{FE0000} 4,51} & {\color[HTML]{FE0000} 0,057}  & {\color[HTML]{FE0000} 0,323} \\
{\color[HTML]{009901} 10} & {\color[HTML]{009901} c1f10}                   & {\color[HTML]{009901} 2,57} & {\color[HTML]{009901} 4,36} & {\color[HTML]{009901} 0,108}  & {\color[HTML]{009901} 0,354} \\
{\color[HTML]{009901} 11} & {\color[HTML]{009901} c1f15}                   & {\color[HTML]{009901} 2,63} & {\color[HTML]{009901} 4,02} & {\color[HTML]{009901} 0,195}  & {\color[HTML]{009901} 0,186} \\
{\color[HTML]{009901} 12} & {\color[HTML]{009901} ff35}                    & {\color[HTML]{009901} 2,56} & {\color[HTML]{009901} 4,02} & {\color[HTML]{009901} 0,172}  & {\color[HTML]{009901} 0,205} \\
{\color[HTML]{009901} 13} & {\color[HTML]{009901} ff50}                    & {\color[HTML]{009901} 2,59} & {\color[HTML]{009901} 4,21} & {\color[HTML]{009901} 0,080}  & {\color[HTML]{009901} 0,428} \\
{\color[HTML]{3166FF} 14} & {\color[HTML]{3166FF} Rhadamanthys Linea}         & {\color[HTML]{3166FF} 3,47} & {\color[HTML]{3166FF} 3,83} & {\color[HTML]{3166FF} 0,728}  & {\color[HTML]{3166FF} 0,011} \\
{\color[HTML]{3166FF} 15} & {\color[HTML]{3166FF} Phaidra Linea}           & {\color[HTML]{3166FF} 3,47} & {\color[HTML]{3166FF} 3,89} & {\color[HTML]{3166FF} 0,693}  & {\color[HTML]{3166FF} 0,0018} \\
16                        & MUD1                                           & 2,65                        & 4,28                        & \multicolumn{1}{l}{-}        & \multicolumn{1}{l}{-}       \\
17                        & MUD2                                           & 3                           & 5,67                        & \multicolumn{1}{l}{-}        & \multicolumn{1}{l}{-}       \\
18                        & Greenland                                      & 3,19                        & 3,95                        & \multicolumn{1}{l}{-}        & \multicolumn{1}{l}{-}  
    \end{tabular}
    \caption{Data for fracture mosaics plotted in Fig. \ref{fig:2}.}
    \label{table:1}
\end{table}

\section*{Planetary Fracture Mosaics}
{\djj{We now examine the geometry of 2D fracture mosaics sampled from planetary bodies across the solar system: Venus, Mars and Europa. A catalog of suitable planetary images was compiled from a variety of online databases and existing literature (Methods). Fractures in each image were manually traced, and $n^*$ and $v^*$ values for each node and cell, respectively, were counted (Fig. \ref{fig:1}; Supplementary Data Section 6). We performed statistical tests to demonstrate the suitability of convex polygons for describing fracture cells, and also the minimum sample size required for computing reliable averages (Methods). Importantly, our methods are insensitive to image distortion and angle; this makes counting cracks particularly useful for analyzing planetary images, which are often not ideal. The resulting raw data for each image are distributions of $n^*$ and $v^*$ (Fig. \ref{fig:1}, Supplementary Data Section 6.), which are used to compute $\bar{n}^{\star}_1$, $\bar{v}^{\star}_1$ and $p$ (Supplementary Data Section 2.). It is important to point out that these data are independent of any model assumption. Reassuringly, we find that all observed mosaics are contained within the region of the Symbolic Plane corresponding to convex mosaics, and are roughly bounded by the three end members (Fig. \ref{fig:2}, Fig. S3, Fig. S13). {\gd{ We also plotted
our data on
the Symbolic Ternary Diagram (Fig \ref{fig:2}, Fig. S13).}} Some error is incurred with this abstraction, which ignores any nodes that are not X, T or Y; however, the error is always very small for the 15 mosaics we examine here (Supplementary Data Sections 4 and 5). }} 


{\djj{Planetary mosaics form three clusters, that each occupy distinct regions in the Symbolic Plane  {\gd{(Fig. \ref{fig:2})}}. Indeed, we observe that each cluster falls into one of three partitions in the Symbolic Ternary Diagram: mosaics dominated by T junctions (I), Y junctions (II) or X junctions (III). To determine whether these clusters represent distinctly different fracture mechanisms, we consider the dynamical view of the Symbolic Plane (Fig \ref{fig:3}). Each cluster corresponds to a distinct quadrant, defined by the parameter lines $\mu =1, m=1.$
The largest cluster plots entirely in {\gd{partition I (Fig. \ref{fig:2}) on the Symbolic Ternary Diagram and quadrant $A$ on the Symbolic Plane, (Fig. \ref{fig:3})}} closest to the Irregular Limit . This cluster includes, for example, martian fractures infilled by precipitated cements \cite{Siebach2014} and fractured blocks on Venus \cite{Byrne2021}. {\djj{These mosaics are diverse in material composition and environment; their geometry, however, indicates that all of them formed predominantly by irreversible binary breakup.}}  
All mosaics in this cluster have regularity $p < 0.5$ and a modal value of $n^*=2$ (Supplementary Data Section 6., Fig S14.); that is, the majority of nodes in each mosaic are T ($\bar{n}^{\star} = 2$) junctions. We interpret these patterns to be mature crack networks with substantial secondary fracture, that have not experienced significant resurfacing or rejuvenation. 
{\djj{Indeed, quadrant $A$}} corresponds to model values $\mu > 1$ and $m<1$ (Fig. \ref{fig:3}), both of which are significantly less than 1, confirming that evolution of mosaics within this quadrant was dominated by binary breakup with little or no healing or node twisting (Fig. S4).

There is a cluster of four mosaics in {\gd{partition II of the Symbolic Ternary Diagram (Fig \ref{fig:2}) and quadrant B of the Symbolic Plane (Fig \ref{fig:3})}} that stand out for being closest to the Voronoi Limit, 
and all of them are on Mars. The closest mosaic to Voronoi is from Amazonis Planitia, and appears to represent polygons formed in a cold polar region where thermal cycling causes ice-rich ground to expand and contract \cite{Mellon2009}. 
The cells in this mosaic are, {\gd{on average,}} remarkably hexagonal ($\bar{v}^{\star} = 5.5$), suggesting that these polygons experienced numerous cycles of volume change -- {\djj{potentially related to the presence of frozen water.}}
Other mosaics include: 
a satellite image of Mawrth Vallis that was identified by previous researchers as putative lacustrine mudcracks containing smectite clays \cite{El-Maarry2014}; an image taken by the Zhurong Rover on a salt-crusted dune in southern Utopia Planitia, where researchers proposed that the observed polygons formed by desiccation or diurnal temperature fluctuations involving brine \cite{Qin2023}; and a satellite image of the Northern Plains on Mars, identified as putative freeze-thaw structures \cite{Jaumann2002}. These four mosaics each have a modal value of {\kr{$n^{\star}=3$}} (Supplementary Data Section 6) -- i.e., nodes are predominantly Y junctions -- making them distinct from all Irregular Mosaics identified earlier. We infer that all martian patterns in quadrant $B$ were formed by numerous cycles of volume expansion and contraction.  {\djj{Indeed, if we reasonably assume that the initial crack networks were finite Gilbert mosaics (Fig. S5)}}, these close neighbors in the Symbolic Plane can all be {\gd{described by evolutionary dynamics controlled}} 
by similar values of $m$ and $\mu$ (Fig. S12) that reflect a predominance of node twisting, and no overprinting. For example, from the modeled trajectory of Mawrth Vallis we infer {\gd{$m=0.256$ and $\mu = 0.251$}} (Fig. \ref{fig:4}). These findings have important implications for the history of water on Mars. The repeated inundation and desiccation implied by the mudcracks, for example, suggests a temperate environment that was proximal to a liquid water source. Hexagonal mudcracks observed by the rover Curiosity in Gale Crater were interpreted to have formed by high-frequency wet-dry cycling \cite{rapin2023sustained}, but that study did not quantify fracture network geometry. Our approach allows us to quantify the significance of cyclic volume change (in terms of node twisting) compared to other fracture processes. Voronoi mosaics are tantalizing targets for future missions seeking out signs of life on Mars.

There are two mosaics {\gd{located in partition III of the Symbolic Ternary Diagram (Fig. \ref{fig:2}) and quadrant $C$ of the Symbolic Plane (Fig. \ref{fig:3})}} that stand apart for their close proximity to the Regular Limit: Rhadamanthis Linea and Phaidra Linea, which are both on Europa. These fracture networks are dominated by X junctions, i.e., cracks that cross other cracks. Few cells appear to have experienced binary fracture, and both mosaics have a modal value of {\kr{$n^{\star}=4$}} (Fig. S16), distinct from all other mosaics we examined. These are the only planetary mosaics examined that have $m >1$, indicating significant overprinting. Recent work by Culberg et al. \cite{Culberg2018} provides a potential explanation: these fractures on Europa are infilled from below by pressurized water, which freezes at the surface. This crack healing would allow for new fractures to propagate across old ones -- a common phenomenon in fractured ice \cite{murdza2022rapid, murdza2023behavior, demmenie2022scratch, colbeck1986theory, schulson2016restoration} -- keeping the mosaic close to the Regular Limit even with significant secondary fracture. Dynamical {\gd{analysis of the}} geometric model supports this interpretation,
indicating that Europa's fracture networks are {\gd{indeed}} mature mosaics with significant secondary fracture ($\mu \gg 1$) and overprinting of cracks ($m > 1$) (Fig. \ref{fig:3}, Fig. \ref{fig:4}; Fig. S12). 
Culberg et al. \cite{Culberg2018} discovered that the double-ridge ice fractures on Rhadamanthys Linea have a similar morphology to ice fractures in Greenland, where this process of healing by infilling was directly observed. The close proximity of these two fracture mosaics, in terms of their network geometry, provides an independent confirmation of their similarity (Fig. \ref{fig:2}). 
Europa is emerging as a prime target in the search for extraterrestrial life, because of the likely subsurface ocean of liquid water \cite{Howell2020}. We suggest that our geometric analysis could be used to pinpoint other fracture networks of interest on various icy worlds.

\begin{figure} [ht]
    \includegraphics[width=12cm]{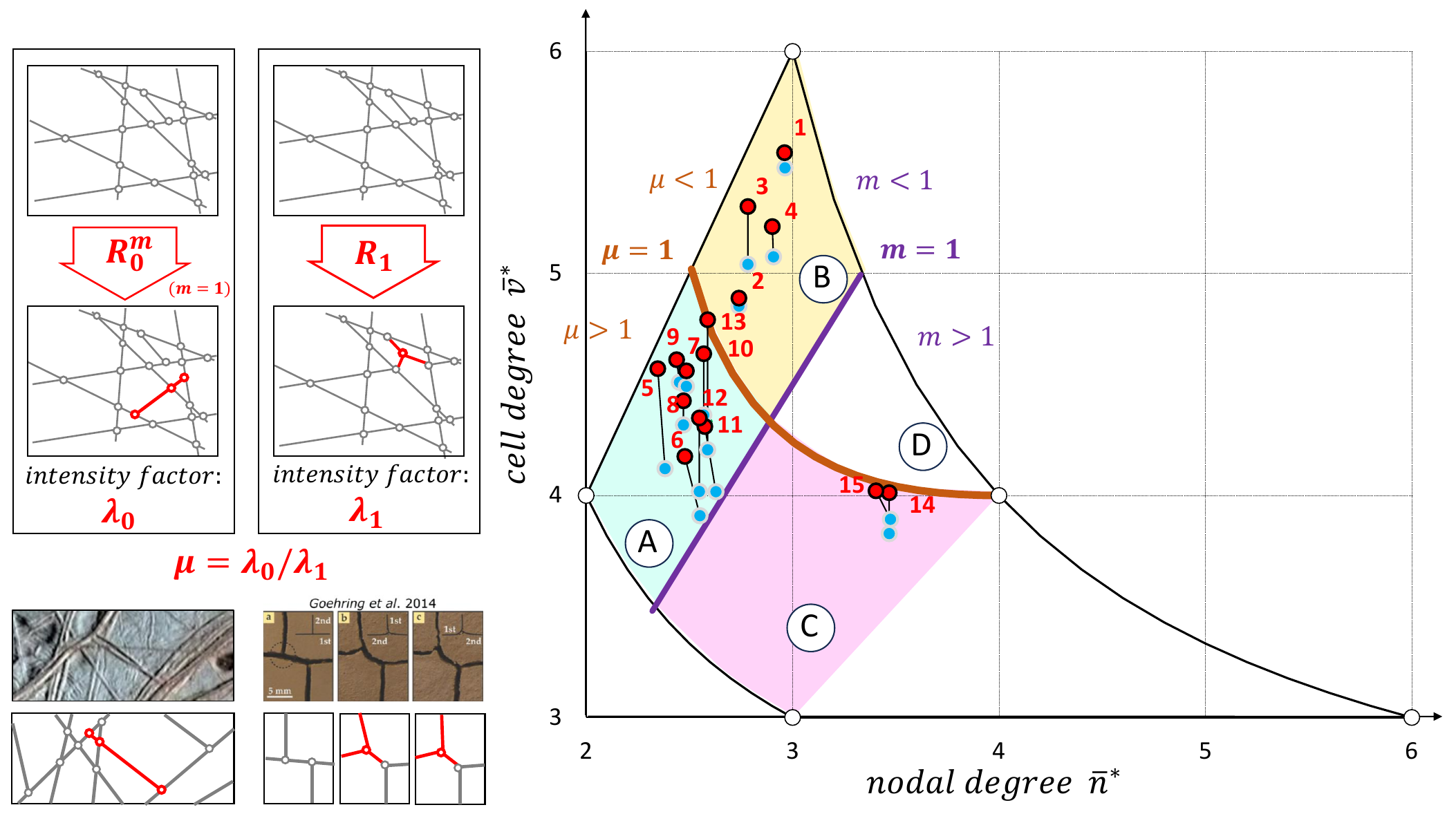}
    \centering
    \caption{\textbf{The dynamical view of the planetary Symbolic Plane.} Upper left: Schematic of the dynamical fracture model showing elementary discrete events $R_0^m$ and $R_1$ with corresponding intensities $\lambda_0, \lambda _1$ driving the evolution. Bottom left: elementary discrete events observed in natural crack patterns and their abstract representation, shown on Europa (left image) and laboratory mudcracks subjected to cyclic volume change \cite{Goehring2010} (right). Right: Symbolic Plane with static estimate $M_1(\bar{n}_1^*,\bar{v}_1^*)$ (blue circle) and dynamical estimate $M_2(\bar{n}_2^*,\bar{v}_2^*)$ (red circle) for each measured mosaic in the Solar System plotted. Numbers refer to the Table \ref{table:1}. Static and dynamic estimates belonging to the same mosaic connected  by thin black lines. Although the position of each mosaic shifts somewhat, no mosaic changes quadrants from the static to dynamical estimate. Quadrants A,B,C,D are delineated by $m = 1$ (purple line) and $\mu = 1$ (brown line). The majority of observed mosaics are in Quadrant A ($m < 1$, $\mu > 1$). Mosaics in Quadrant B ($m < 1$, $\mu <1 1$) are only found on Mars, in materials suspected to experience cyclic volume change. Mosaics in Quadrant C ($m > 1$, $\mu > 1$) appear only on Europa, likely due to crack healing. Quadrant D ($m > 1$, $\mu < 1$) is empty; node twisting and healing do not appear to coexist. }
    \label{fig:3}
\end{figure}

\begin{figure} [ht]
    \includegraphics[width=12cm]{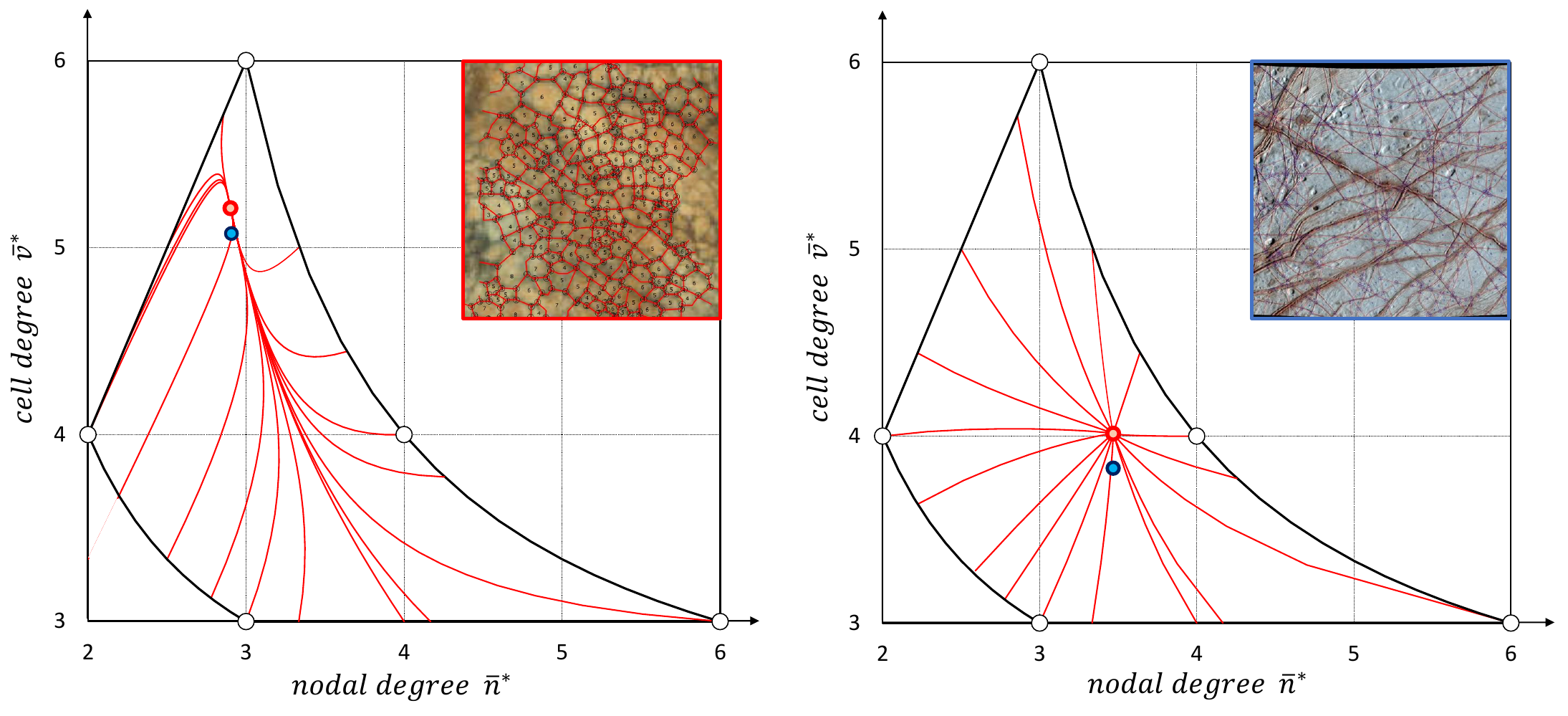}
    \centering
     \caption{\textbf{Evolution of fracture mosaics to maturity.} Example evolutionary trajectories are shown for two systems. Left: Mawrth Vallis mosaic from Mars (example from quadrant B); static estimate $M_1(\bar{n}_1^*,\bar{v}_1^*)=(2.907, 5.075)$ (blue circle), dynamic estimate $M_2(\bar{n}_2^*,\bar{v}_2^*)=(2.90, 5.21)$ (red circle), system parameters $(\mu,m)=(0.251,0.256)$.  Right: Rhadamanthis Linea mosaic from Europa (example from quadrant C); static estimate $M_1(\bar{n}_1^*,\bar{v}_1^*)=(3.46, 3.83)$ (blue circle), dynamic estimate $M_2(\bar{n}_2^*,\bar{v}_2^*)=(3.46, 4.01)$ (red circle), system parameters $(\mu,m)=(3.67,5.36)$. Both examples demonstrate that the dynamical fracture model evolves toward a fixed point for a given parameter set -- regardless of initial conditions -- and that observed mosaics are close to the predicted fixed point, consistent with the Maturity Assumption.
 }\label{fig:4}
\end{figure}

\section*{Discussion}
The observed clustering of fracture mosaics in the Symbolic Ternary Diagram (Fig. \ref{fig:2}), and the formative processes we infer, are supported by the geometric model (Figs. \ref{fig:3}, \ref{fig:4}) -- giving us confidence that our approach has utility for application in planetary exploration.
Most of the planetary fracture mosaics we examined lie closest to the Irregular Limit ({\djj{quadrant $A$}}) in the Symbolic Plane, delineated by $\mu > 1$ and $m<1$ (Fig. \ref{fig:3}), indicating that they formed predominantly by irreversible binary breakup. {\djj{The prevalence of hierarchical fracture patterns throughout the solar system demonstrates the propensity for thin cracked shells to accumulate damage with limited resurfacing or rejuvenation, regardless of material composition}}. Exceptions to this trend reveal systems in which healing or cyclic volume change are likely involved. It is intriguing that outliers in the Symbolic Ternary Diagram correspond to patterns that likely involved liquid water. 

Our interpretations of the influence of water on planetary fracture geometry would be bolstered with a more populated Symbolic Ternary Diagram. One bottleneck is that manual extraction of crack networks from images is time consuming; we were unable to automate the process using existing network extraction tools \cite{Dirnberger2015}. This limitation may soon be overcome with rapidly advancing machine-learning techniques \cite{lai2020vulnerability, Kirillov}. More restrictive is the manual process for discovering putative fracture networks, which requires combing through planetary image archives. A related challenge is that fracture mosaics can be geometrically similar to convection patterns \cite{Thiele1998}, and we don't know of any approach (yet) that can discriminate between them from geometry alone. For example, polygonal ground on Sputnik Planum (Pluto) was first proposed to have formed by fracture of the surface ice \cite{Stern2015}, but more recently researchers have argued in favor of convection of the underlying ice \cite{McKinnon2016, Trowbridge2016}. Pluto's polygonal ground can certainly be plotted on the Symbolic Plane 
but any inference of fracture process is uncertain without ruling out alternative processes like convection. When emerging techniques for network extraction can be applied successfully to automate detection and analysis of fracture networks, however, we will have an even more powerful tool for discovering potentially habitable landscapes on planetary bodies. 

\section*{Methods}
\subsection*{Planetary Image Analysis}
\noindent Planetary images are acquired from existing publications and online databases, specifically the NASA Photo Journal and the NASA Planetary Data System (PDS) Image Atlas, both of which are managed by the Jet Propulsion Laboratory (JPL). Skeletonization is done by hand in Inkscape by tracing the fractures visible in the image (Fig. \ref{fig:1}). From there, a series of quality assurance tests are conducted to ensure that the network is well-represented by convex polygons, that the impact of edge-effects is minimal, and that the averages are reliable and statistically significant. Convex polygon approximation is conducted by calculating the difference in area when drawing straight lines connecting nodes in order to make all polygons regular (Fig. S2). No mosaic had an error $>10\%$. Following skeletonization, values of $n^*$ and $v^*$ are assigned to each node and cell, respectively, in the mosaic (Fig. \ref{fig:1}).

Once a network is judged suitable based on the criteria established above, the values of $n*$ and $v^*$ are counted and their distributions plotted as histograms. 
The average nodal and cell degrees are then plotted as a pair in our Symbolic Plane, which encompasses all possible $\bar{n}^*$ and $\bar{v}^*$ values for a mosaic composed of 2D convex polygons. {\gd{We also count the number of T, Y and X nodes to determine the respective relative frequencies $\hat T, \hat Y$ and $\hat X$, and plot the corresponding
point on the Symbolic Ternary Diagram.}}  There is occasional ambiguity in the assignment of $n^*$ and $v^*$ values {\gd{and the classification of nodes}}; for example, whether a node is a T or a Y. These special cases are recorded in a running database in order to maintain internal consistency. For all networks, the number of ambiguous cases are a very small fraction of the overall counts; the network with the greatest number of fringe cases, Rhadamanthys Linea, has 6 out of a total of 541 nodes, making the percent of fringe cases about 1\%. 

\subsection*{Geometric Evolution Model}
\noindent  Our model is a modest generalization of the theory presented in \cite{Balint2023} where the continuous time evolution equations of crack mosaics, driven by two discrete local events, are derived.
This theory admits the study of trajectories in the Symbolic Plane {\gd{and the Symbolic Ternary Diagram}} along which crack mosaics {\gd{$M$}} move, until they reach their attractor which appears as a fixed point $F{\gd{(M)}}$ in the Symbolic Plane and as a fixed point {\gd{and $G(M)$}} in the Symbolic Ternary Diagram.
Our theory for the evolution is identical to  \cite{Balint2023} except for one modification; for the discrete step representing secondary cracks, we allow those cracks to traverse several cells (rather than always terminate at {\gd{the next}} cell boundary).
Plotting crack mosaics on the Symbolic Plane permits comparison {\gd{based on combinatorial averages;}} 
however, close neighbours on the Symbolic Plane may {\gd{exhibit radically different geometric features.}}
The dynamical {\gd{evolution}} model
facilitates a deeper, physically meaningful comparison between crack mosaics.
{\gd{We start with the assumption that the current location in the Symbolic Ternary Diagram is identical to the attractor $G(M)$.
This \emph{Maturity Assumption} is natural; newly-formed mosaics move rapidly along their trajectory in the Symbolic Plane, but their evolutionary speed diminishes toward zero as they approach the fixed point (Supplementary Information, Section 4.4 and 5.4). As a consequence, it is far more likely that at any random time, any observed mosaic $M$ is mature and close to its fixed point, than that it is young and rapidly evolving. Note that the dynamical model predicts the distribution of nodal degrees but not cell degrees. Since the Symbolic Ternary Diagram only uses the former, we expect that the estimate $G(M)$ on the attractor will be much
more efficient than that on the Symbolic Plane, where the unpredictable variation of cell degrees can cause large fluctuations. Using the Maturity Assumption, we can find the point $M_2$ on the Symbolic Plane that represents the (estimated) attractor with equation (\ref{eq:dynamic}). Using the dynamical model, from the location of the attractor we can compute estimates for the system parameters $\mu$ and $m$. These parameters, coupled with 
the differential equation governing evolution, provide all possible trajectories (see Supplementary Equations 22-23.}}


{\gd{In the dynamical model all newly created nodes are either T,Y or X type. This implies, via the Maturity Assumption, that the ratio of non-TYX nodes should be small for all mosaics. Indeed, for the 6 ``outlier'' mosaics of interest (quadrants B and C in Fig. 3) we found that their ratio was below 3\% in all cases, confirming the Maturity Assumption. The same assumption implies that if the system parameters $m,\mu$
remain constant then the values $\hat T, \hat Y, \hat X$ shall also remain constant.
To further check the veracity of our assumptions, for each investigated mosaic $M$ we identify a generating trajectory $t(M)$ that (i) connects the current state $M_1$ of the mosaic with its estimated attractor $M_2$ in \emph{forward time}, and (ii) connects $M_1$ with its estimated initial condition $M_0$ (located among reasonable initial conditions) in \emph{backward time}. We were able to identify a generating trajectory for all 6 investigated mosaics (see SI, Section 4 for details) confirming the
validity of the model assumptions.}}

\section*{Acknowledgements}
Research was supported by: NASA through the PSTAR program (Grant 80NSSC22K1313) to D.J.J; the NKFIH Hungarian Research Fund (Advanced Grant 149429), and a grant from BME FIKP-VÍZ by EMMI, to G.D; the EKÖP-24-3 program funded by ITM and NKFI, the Doctoral Excellence Fellowship Programme (DCEP) funded by ITM and NKFI, the Budapest University of Technology and Economics, and the gift from the Albrecht Science Fellowship, to K.R.

\section*{Author contributions}
S.S. initiated the study, and collected and analyzed fracture mosaic data. 
K.R. analyzed fracture mosaic data and developed theory.
D.J.J. supervised the research.
G.D. supervised the research, developed theory and performed numerical simulations.All authors contributed to writing the manuscript.

\section*{Competing interests}
The authors declare that we have no competing interest.

\section*{Materials \& Correspondence}
Correspondence should be addressed to G.D. (domokos@iit.bme.hu) or D.J.J. (sediment@sas.upenn.edu).

\bibliographystyle{unsrt}
\bibliography{bib}

\end{document}